\documentstyle[pre,aps,multicol,epsf]{revtex}

\begin{document}
\draft

\title{Two-Dimensional Polymers with Random Short-Range Interactions}
\author{Ido Golding and Yacov Kantor}
\address{School of Physics and Astronomy, Tel Aviv University,
Tel Aviv 69 978, Israel}
\maketitle

\begin{abstract}
We use complete enumeration and Monte Carlo techniques to study two-dimensional self-avoiding polymer chains with quenched ``charges'' $\pm 1$. The interaction of charges at neighboring lattice sites is described by $q_i q_j$. We find that a polymer undergoes a collapse transition at a temperature $T_{\theta}$, which decreases with increasing imbalance between charges. At the transition point, the dependence of the radius of gyration of the polymer on the number of monomers is characterized by an exponent $\nu_{\theta} = 0.60 \pm 0.02$, which is slightly larger than the similar exponent for homopolymers. We find no evidence of freezing at low temperatures.
\end{abstract}

%-----------------------------------------------------------
% Document
%-----------------------------------------------------------

%\epsfverbosetrue   % This is to know the size of the figure

\begin{multicols}{2}
\narrowtext

A polymer in a solvent is subject to monomer-monomer interactions which consist of a short-range repulsion and a slightly longer range attraction. At high temperatures (or good solvent conditions), the repulsive interactions are dominant, and the polymer is swollen: Its radius of gyration $R_g$ scales with the number of monomers $N$ as $R_g \sim N^{\nu}$, with $\nu > 1/2$. As temperature $T$ is lowered (or solvent conditions worsen), a point is reached, called the $\theta$-point, where the repulsive and attractive interactions effectively cancel and the polymer scales, for space dimensions $d \geq 3$, like an ideal random walk ($\nu = 0.5$) \cite{florybook,degennes55}. For $T$ smaller than the transition temperature $T_{\theta}$, the attractive interactions prevail and the polymer collapses into a compact shape, with $\nu = 1/d$.
Such {\em homopolymers} are often modeled by self-avoiding walks (SAWs) on a discrete lattice, with the attractive interactions included by introducing a negative energy for each pair of monomers residing on neighboring lattice sites. Numerous Monte Carlo (MC) and exact enumeration studies of the $\theta$-point have been performed \cite{kremer2879,baumgartner1407,ishinabe3181}.

The collapse transition of {\em heteropolymers} is particularly interesting in view of its possible relation to the problem of protein folding \cite{chan24,creightonbook}. While models based on random heteropolymers significantly over-simplify the complexity of real proteins, they bring in fresh perspectives from the statistical mechanics of random systems and spin glasses \cite{bryngelson7524,shakhnovich187}. 
A question of high interest (with no definite answer) is whether the interactions between different monomer types can modify the collapse transition of a polymer \cite{harris347}. 
Another interesting feature of heteropolymer chains concerns the compact state: Is there a freezing transition at some temperature $T_f ( < T_{\theta})$, below which the configurational entropy per monomer vanishes, and few conformations with low energy dominate \cite{shakhnovich187,shakhnovich5967}. This freezing transition is assumed to be analogous to the glass transition in Derrida's random energy model (REM) \cite{derrida2613}, although this analogy is not always valid \cite{pande3987}.

\begin{figure}[]
\epsfysize=13\baselineskip
\centerline{\hbox{
      \epsffile{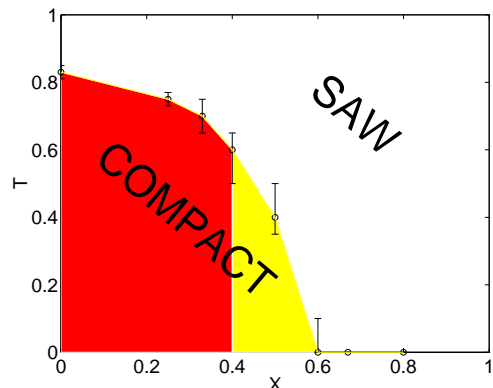}  }}
\caption [A Picture]
         {\protect\footnotesize Phase diagram of a random heteropolymer in the plane of temperature (T) and excess charge (X). Vertical bars indicate estimated uncertainties in $T_{\theta}$. The lighter shaded area indicates $X$ values where the results are rather ambiguous, due to the poor quality of the MC results.}
\label{phase}
\end{figure}

We study self-avoiding polymer chains on a two-dimensional (2D) square lattice. Each polymer chain is formed from two types of monomers, with charges $q_i = \pm 1$. The interaction Hamiltonian is:

\begin{equation}
\label{ourmodel}
H_I = \frac{1}{2}\sum_{i,j}{{q_i q_j {\Delta}_{ij}}}\;,
\end{equation} 
where  ${\Delta}_{ij} = 1$ if monomers $i$ and $j$ are located on adjacent lattice sites and ${\Delta}_{ij} = 0$ otherwise. The homogeneous repulsion is imposed by the constraint of self-avoidance.
The use of a 2D system enables us to investigate longer chains than in 3D models \cite{kantorpp}, and thus examine aspects not studied before.
We find that a polymer undergoes a $\theta$-transition, with a critical exponent $\nu_{\theta} = 0.60 \pm 0.02$, which is slightly larger than the value for the homopolymer case, meaning that heterogeneity is a significant perturbation in this model. 
For a polymer with unbalanced  numbers of positive and negative charges $(N_{+} \neq N_{-})$ 
we find that $T_{\theta}$ decreases, and eventually vanishes, with increasing the excess charges fraction $X \equiv \; \mid \! N_{+} - N_{-} \! \mid \!/ N$ (see Figure \ref{phase}).
Finally, we explore the ground state and the energy landscape of a neutral (i.e.\ $X=0$) polymer, in an attempt to find evidence for the existence of a glass-like freezing transition for such a polymer. We do not find much evidence for this transition, a result which is in accordance with theoretical predictions for our model. 

Following Kantor and Kardar \cite{kantorpp}, we study polymers with quenched heterogeneity and a fixed number of monomers $N$.
The calculation of thermodynamic quantities is performed as follows: For chain lengths of up to 15 steps (16 monomers), we completely enumerate all possible spatial conformations, and average over all possible charge sequences (quenches) for a given $X$. The number of the spatial conformations grows exponentially with $N$ \cite{degennesbook}. However, we take advantage of lattice symmetries to reduce the number of independent configurations. Thus, for a 16-monomer neutral chain we have to enumerate 802,075 different configurations (unrelated by symmetry), and average further over 12,870 quenches, giving $\sim 10^{10}$ possibilities.
For chain lengths of up to 23 steps, we still enumerate all conformations, but average over a limited number (20-100) of quenches. For chains of up to $N=100$, we use MC simulations, applying the ``pivot'' algorithm \cite{kremerpp}. 

\begin{figure}[]
\epsfysize=15\baselineskip
\centerline{
        \hbox{\epsffile{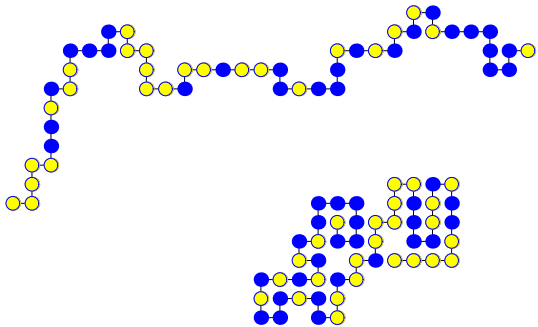}  }
        }
\caption [A Picture]
         {\protect\footnotesize Expanded and collapsed configurations of a neutral 50-monomer polymer at $T=2$ (top) and $T=0.6$ (bottom), obtained by MC simulation. Oppositely charged monomers are denoted by dark and light filled circles.}
\label{conf}
\end{figure}

\begin{figure}[]
\epsfysize=13\baselineskip
\centerline{
        \hbox{\epsffile{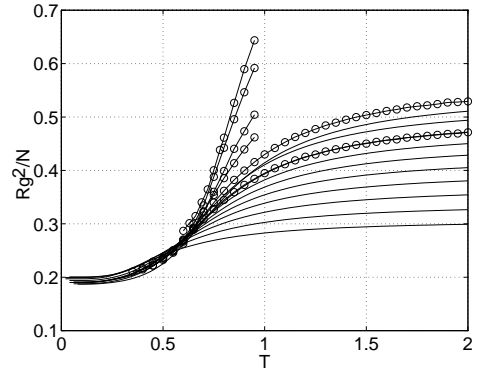}  }
        }
\caption [A Picture]
         {\protect\footnotesize Squared radius of gyration, divided by chain length, for neutral polymers. Distances are measured in lattice constants, temperature is normalized by monomer interaction energy. Curves are for all quenches (unless stated otherwise) and the following number of monomers (from bottom right) : 6 , 8 , 10 , 12 , 14 , 16 , 18 (100 quenches) , 20 (100 quenches enumeration, 50 quenches MC) , 22 (100 quenches) , 24 (20 quenches) , 26 (50 quenches), 36 (25 quenches) , 50 (10 quenches) , 80 (10 quenches) , 100 (10 quenches). Solid lines represent results of enumeration, connected circles represent results of Monte Carlo simulation.}
\label{rg2}
\end{figure}

\begin{figure}[]
\epsfysize=13\baselineskip
\centerline{
        \hbox{\epsffile{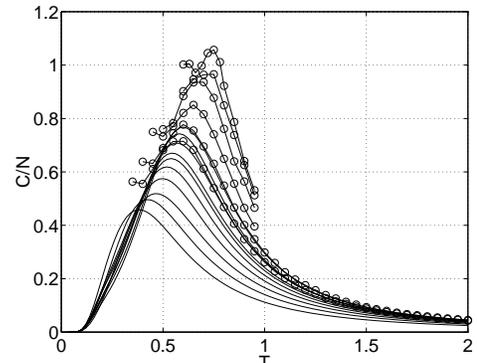}  }
        }
\caption [A Picture]
         {\protect\footnotesize Heat capacity divided by chain length, for neutral polymers. For details of chains see Fig.\ \ref{rg2}.}
\label{c}
\end{figure}

The presence of a collapse can be clearly seen in Fig.\ \ref{conf}, depicting two conformations of a 50-monomer neutral polymer, sampled by the MC procedure at high and low temperature.
We expect the polymer to be SAW-like (with $\nu_{\rm SAW}=0.75$) at high temperature, and become compact ($\nu_{\rm compact}=1/d=0.5$) when we lower the temperature. Thus, $R_g^2/N \sim {\rm const}$ in the collapsed state and $R_g^2/N \sim N^{0.5}$ in the swollen state. Temperature dependence of $R_g^2/N$ for various $N$s, depicted in Fig.\ \ref{rg2}, indeed exhibits such behavior.
The presence of a phase transition is also indicated by the peak in the heat capacity, as shown in Fig.\ \ref{c}. The peak grows and slightly shifts toward higher temperatures with increasing $N$. Out results bear qualitative resemblance with the known behavior of homopolymers (see, e.g. \cite{baumgartner1407}).
 

We cannot use the simple method used to find  $T_{\theta}$ in the 3D case \cite{kantorpp}, which is to observe the intersection of the graphs of $R_g^2/N$ vs. $T$ for different $N$s (using the facts that $ \nu_{\theta} = \nu_{\rm ideal} = 0.5$ and $\nu_{\rm SAW} > \nu_{\theta} > \nu_{\rm compact}$). This is because in the 2D case, $\nu_{\rm compact} = \nu_{\rm ideal} = 0.5$. Moreover, unlike in 3D, it is believed that for 2D, $\nu_{\theta} > \nu_{\rm ideal}$ \cite{degennes55,vilanove1502}, and the 2D value of $\nu_{\theta}$ is not known a priori.
The method we use to estimate $T_{\theta}$ and $\nu_{\theta}$ is based on the observation, made by de Gennes, that the ${\theta}$-point is a {\em tricritical} point \cite{degennes55,marqusee7159}. Based on the general theory for tricritical phase transitions, it has been shown theoretically \cite{degennes55,daoud973} and verified numerically for homopolymers \cite{kremer2879,baumgartner1407}  that $R_g$ for different temperatures and chain lengths can be described using a single scaling function:

\begin{equation}
\label{scaling}
R_g^2 / N^{2 \nu_{\theta}} = f_{\pm}(N^{\phi}\tau)\;,
\end{equation}
for $N \gg 1$ and $\tau \ll 1$, where $\tau \equiv |T-T_{\theta}|/T_{\theta}$ and $\phi$ is a crossover exponent. The scaling function $f_{\pm}(x)$ should have the following limits:

\begin{equation}
\label{scaling2}
f_{\pm}(x) \propto \left\{ \begin{array}{ll}
                {\rm const}             & {\rm for} \;\mbox {$x \rightarrow 0$}\\ 
                x^{2(\nu_{\rm SAW}-\nu_{\theta})/\phi}          & {\rm for}\;\mbox {$x \rightarrow \infty \;,\; T>T_{\theta} $}\\
                x^{2(\nu_{\rm compact}-\nu_{\theta})/\phi}      & {\rm for}\;\mbox {$x \rightarrow \infty \;,\; T<T_{\theta} $}\;.\\
                 \end{array}
        \right . 
\end{equation}
It is easily seen that the asymptotes of $f_{\pm}(x)$ at high and low temperatures reconstitute the behavior of a SAW and of a compact chain, respectively.

\begin{figure}[]
\epsfysize=13\baselineskip
\centerline{
        \hbox{\epsffile{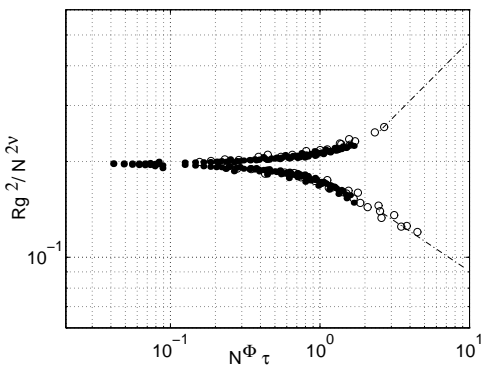}  }
        }
\caption [A Picture]
         {\protect\footnotesize Logarithmic plot of $R_g^2/N^{2\nu_{\theta}}$ vs. $N^{\phi} \tau$, close to the tricritical temperature. Data from chain lengths 7--99 is used, with values of $\tau$ up to 0.25. Parameters used: $\nu_{\theta}=0.60 ,\; T_{\theta}=0.83 ,\; \phi=0.636$. Dots denote enumeration data, circles denote MC data. Dashed lines show theoretical asymptotes (Eq. \ref{scaling2}), with amplitude fitted to data .}
\label{scale}
\end{figure}

\begin{figure}[]
\epsfysize=13\baselineskip
\centerline{\hbox{
      \epsffile{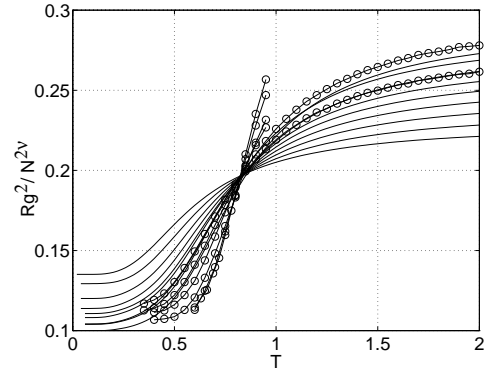}  }}
% parameters of fig8
\caption [A Picture]
         {\protect\footnotesize $R_g^2 / N^{2\nu_{\theta}}$ vs. temperature, for neutral polymers of length 7-99 (For details see Fig.\ \ref{rg2}). The graphs intersect at the $\theta$-point.}
\label{intersect}
\end{figure}

Fig.\ \ref{scale} depicts the scaling function $R_g^2/N^{2 \nu_{\theta}}$ vs. the scaling variable $N^{\phi} \tau$, using data of all chain lengths (7--99) for temperatures satisfying $\tau \leq 0.25$, and choosing the parameters $T_{\theta}$ , $\nu_{\theta}$ and $\phi$ so that the points fall on two converging lines as required. As seen in the figure, a very good data collapse is achieved for $T_{\theta} = 0.83$ and $\nu_{\theta} = 0.60$, and the resulting lines also approach the slopes of the theoretical asymptotes. It should be further pointed out that the scaling behavior is found to be very sensitive to the value of $\nu_{\theta}$, slightly less to the value of $T_{\theta}$, and quite insensitive to the choice of $\phi$. 
After examining numerous figures of the like of Fig.\ \ref{scale}, using various parameter values, we evaluate the transition parameters to be
$T_{\theta} = 0.83 \pm 0.02$ and
$\nu_{\theta} = 0.60 \pm 0.02$,    
where the errors were estimated according to the parameter values where data did not collapse anymore (according to our subjective judgment). 
We verify these values by plotting $R_g^2/N^{2 \nu_{\theta}}$ vs. $T$ for various chain lengths (Fig.\ \ref{intersect}). The curves should intersect at $T=T_{\theta}$ (because $\nu_{\rm SAW} > \nu_{\theta} > \nu_{\rm compact}$). This is indeed what happens, within the estimated errors of the parameters and of the simulated data.

Our value of $\nu_{\theta}$ seems to be larger than the 2D homopolymer value, for which most estimates lie in range 0.51-0.58 \cite{vilanove1502}. In order to support this claim,  we have used our own enumeration and MC procedures (properly altered) to simulate homopolymers, and found that $\nu_{\theta}^{\rm homo} = 0.55 \pm 0.01$. Therefore, our conclusion is that $\nu_{\theta}^{\rm hetero} > \nu_{\theta}^{\rm homo}$. 
Simple dimensional arguments show that randomness is marginally relevant for ideal polymers in 2D. One might expect that swelling of the chain at the $\theta$-point will render randomness irrelevant. Beyond simple arguments \cite{harris347} no extensive analysis of the problem has been performed.

What happens to the collapse transition if the charges on the chain are not exactly balanced?
We have repeated the procedures described above for increasing values of excess charge $X$. A decrease in $T_{\theta}$ is observed with increasing $X$, up to a value $X \approx 0.6$ where the collapse vanishes and SAW behavior prevails at all temperatures. The phase diagram in $(X,T)$ plane (Fig.\ \ref{phase}) is similar to that obtained for 3D \cite{kantorpp}. 

Finally, we have attempted to find evidence of a freezing transition for a neutral polymer, analogous to the glass-like transition occurring in the random energy model, which was observed for other heteropolymer models \cite{shakhnovich5967}. 
An indicator of the transition \cite{shakhnovich5967} is the parameter $x(T) \equiv 1 - \sum_k {p_k^2}$, where $p_k$ is the normalized Boltzmann weight of a given conformation $k$. $x(T)$ characterizes the number of conformations which are thermodynamically relevant at a given temperature, and in the case of a freezing transition it should decrease -- when averaged over all quenches -- from $x \approx 1$ at the freezing temperature $T_f$ to $x \approx 0$ at $T=0$. Investigation of our model, however, shows that $x(T)$ does not reach zero value even at $T=0$, due to the degeneracy of the ground state. In addition, the decrease in $x$ occurs in the same temperature range where the polymer collapses, meaning that the decrease in number of relevant conformations comes mainly from the folding -- as is the case of homopolymers.
Another attempt to validate the analogy with the REM was done by examining the similarity between the degenerate ground state conformations of a randomly chosen polymer. The similarity between each two configurations was characterized by the number of monomer pairs which are nearest-neighbors in both of the configurations \cite{shakhnovich5967}. It was found that, on the average, there is indeed a significant dissimilarity between ground state conformations. However, this dissimilarity is not pronounced enough to declare that the ground states are definitely ``structurally different'', as is the case in the REM.

We note, that the seeming absence of a freezing transition in our model may come from specific features of this model: The lack of a strong attractive homopolymeric term in the Hamiltonian, which would lead to the formation of a ``molten globule'', and a lack of sufficient heterogeneity in the monomer-monomer interaction. These two may be required in order to achieve an analogy between a heteropolymer and the REM, and are usually included in model studies, both analytical and numerical, which exhibit a  freezing transition \cite{stafos2898}. 
In addition, it appears \cite{creightonbook,shakhnovich1647} that formation of a unique structure in heteropolymers is very sensitive to space dimensionality, with $d=2$ being a marginal and nonuniversal case that strongly depends upon the type of lattice, type of interaction and so on.

In conclusion, we have investigated a 2D lattice model of polymers, with a quenched random short-range interaction. We have seen that a neutral polymer undergoes a tricritical $\theta$-transition with a critical exponent  $\nu_{\theta} = 0.60 \pm 0.02$, a value that seems to be higher than the homopolymer value. 
Since the difference is rather small, further MC studies with larger $N$, and renormalization-group studies, are needed to verify the difference between the exponents.
For non-neutral polymers, we have observed a decrease in the $\theta$-temperature with increasing excess charge, until the collapse disappears and SAW behavior prevails at all temperatures.
Finally, we did not find evidence for the existence of a glass-like freezing transition for a neutral polymer, a result which seems to be in accordance with theoretical predictions for this model. 

Acknowledgments: IG would like to thank E.\ Brenner for useful
discussions. This work was supported by the US-Israel BSF under grant No.\ 92-26.

%-----------------------------------------------------------
% References
%-----------------------------------------------------------
% \pagebreak

\end{multicols}
\end{document}